\documentclass[structabstract]{aa}
\usepackage{graphicx}
\usepackage{txfonts}
\usepackage{longtable}
\usepackage{natbib}
\bibpunct{(}{)}{;}{a}{}{,}
\begin{document}

\title{On the possible common origin of M16 and M17
\thanks{Based on observations obtained at the Centro Astron\'omico Hispano-Alem\'an, Calar Alto, Spain}
}
\author{F. Comer\'on\inst{1}
\and J. Torra\inst{2}
}
 \institute{
  European Southern Observatory, Karl-Schwarzschild-Str. 2, D-85748 Garching bei M\"unchen, Germany\\
  \email{fcomeron@eso.org}
  \and
  Institut de Ci\`encies del Cosmos (ICCUB-IEEC), Barcelona, E-08028, Spain
  }
%
%\offprints{}
%\mail{}
%\titlerunning{}
%\authorrunning{}
%\date{\today}
%
\date{Received; accepted}
\abstract
% context heading (optional)
{It has been suggested that the well-studied giant HII regions M16 and M17 may have had a common origin, being an example of large-scale triggered star formation. While some features of the distribution of the interstellar medium in the region support this interpretation, no definitive detection of an earlier population of massive stars responsible for the triggering has been made thus far.}
% aims heading (mandatory)
{We have carried out observations looking for red supergiants in the area covered by a giant shell seen in HI and CO centered on galactic coordinates $l \sim 14^\circ 5$, $b\sim +1^\circ$ that peaks near the same radial velocity as the bulk of the emission from both giant HII regions, which are located along the shell. Red supergiants have ages in the range expected for the parent association whose most massive members could have triggered the formation of the shell and of the giant HII regions along its rim.}
% methods heading (mandatory)
{We have obtained spectroscopy in the visible of a sample of red stars selected on the basis of their infrared colors, whose magnitudes are consistent with them being red supergiants if they are located at the distance of M16 and M17. Spectroscopy is needed to distinguish red supergiants from AGB stars and RGB stars, which are expected to be abundant along the line of sight.}
% results heading (mandatory)
{Out of a sample of 37 bright red stars, we identify four red supergiants that confirm the existence of massive stars in the age range between $\sim 10$ and $\sim 30$~Myr in the area. At least three of them have Gaia DR2 parallaxes consistent with them being at the same distance as M16 and M17.}
% conclusions (optional)
{The evidence of past massive star formation within the area of the gaseous shell lends support to the idea that it was formed by the combined action of stellar winds and ionizing radiation of the precursors of the current red supergiants. These could be the remnants of a richer population, whose most massive members have exploded already as core-collapse supernovae. The expansion of the shell against the surrounding medium, perhaps combined with the overrun of preexisting clouds, is thus a plausible trigger of the formation of a second generation of stars currently responsible for the ionization of M16 and M17.}

%\abstract {}


\maketitle
%________________________________________________________________

\section{Introduction} \label{intro}

The close proximity in the sky of M16 and M17, two of the nearest giant HII regions of our galactic neighbourhood lying at a similar distance from the Sun, naturally leads to the question of whether they are physically related, and whether they may share a common origin \citep{Moriguchi02,Oliveira08, Nishimura17}. Both giant HII regions are projected on the contour of a giant bubble-shape structure, outlined in the distribution of HI and CO emission as first noted by \citet{Moriguchi02}. This suggests that the formation of M16 and M17 could have been triggered by the expansion of the bubble, powered by a previous generation of massive stars near its center, thus representing an example of triggered star formation at the scale of several tens of parsecs \citep{Elmegreen98}. Given the ages of the giant HII regions, the timescale of expansion of a wind-blown bubble, and the short lifetimes of massive stars, it is to be expected that the most massive members of that previous generation may have exploded as supernovae several Myr ago. The spatial dispersion of the members of the association that must have taken place progressively during its existence, combined with the distance of 2~kpc to the M16/M17 complex and the large amount of unrelated foreground and background stars in that general direction, would make it extremely difficult to identify even its currently hottest members still remaining on the main sequence. \citet{Moriguchi02} noted the presence of O and early B stars in the area and proposed them to be part of a massive star population responsible for having caused the bubble, but a review of their properties shows them to be generally too bright to be at the distance of the bubble and the giant HII regions, most likely being instead members of a foreground population.

Red supergiants stars offer a venue to circumvent this problem. They are the descendants of
stars with initial masses between 7 and 40~M$_\odot$ \citep{Hirschi10}, which reach this phase after leaving the main sequence at ages ranging from 5~Myr to 50~Myr, the precise value depending on both the initial mass and the initial rotation velocity \citep{Ekstroem12}. The duration of the red supergiant phase ranges from several hundred thousands of years to about 2~Myr. Although relatively short-lived, red supergiants are found in significant numbers in rich stellar aggregates with ages extending up to a few tens of Myr, and their luminosity makes them stand out at near infrared wavelengths, therefore making them suitable and easily accessible probes of past massive star formation in regions where O-type stars have already completed their lifecycles \citep{Davies07,Clark09,Messineo14,Negueruela16,Alegria18}.

Here we present the results of a search for the vestiges of the massive stellar association that could have triggered the formation of the shell containing M16 and M17 through the identification of red supergiants in the area. We present our target selection criteria and our findings, through which we produce a crude estimate of the overall massive star content of the association and the mechanical energy deposited in the bubble, which suggest that the scenario of triggered formation of M16 and M17 is indeed plausible.

\section{Target selection\label{targets}}

M16 and M17 are giant HII regions located at similar distances from the Sun. The distance to M16 has been estimated to be 1.8~kpc \citep[][and references therein]{Oliveira08}, while that of M17 has been determined to be $1.98^{+0.16}_{-0.12}$ \citep{Xu11} through the VLBI trigonometric parallax of masers associated with its massive star formation sites. The ages have been estimated at less than 3~Myr for M16 \citep[e.g.][]{Guarcello12} and 1~Myr or less for M17 \citep[e.g.][]{Hanson97}, although evidence for an older population in the former region has been reported by \citet{Demarchi13}. The average radial velocities of the ionized gas in both regions are similar, $v_{\rm LSR} = +26.3$~km~s$^{-1}$ for M16 and $v_{\rm LSR} = +16.8$~km~s$^{-1}$ for M17 \citep{Lockman89}. Extensive reviews on each of these regions can be found in \citet{Oliveira08} and \citet{Chini08}.

The large shell-like disturbance noted by \citet{Moriguchi02} is well seen in the data extracted from the galactic HI atlas of \citet{Hartmann97} and the galactic disk CO atlas of \citet{Dame01} at radial velocities $+15 \lesssim v_{\rm LSR} {\rm (km \ s^{-1})} \lesssim +20$ between galactic longitudes $12^\circ  \lesssim l \lesssim 18^\circ$, centered near $l \sim 14^\circ 5$, $b\sim +1^\circ$. Both M16 and M17 lie along the rim of the shell, which is incomplete at high galactic longitudes, perhaps indicating that it has formed a chimney-like structure venting gas to the galactic halo. There are no known star clusters possibly associated with the bubble. Of the four star clusters projected on the area listed by \citet{Dias02} and the successive updates of their calalog, three have estimated ages in excess of 250~Myr. The fourth one, NGC 6561, is listed as having an age of 8~Myr, but at 3.4~kpc it is too distant to be physically associated with the region under study.

We used the 2MASS All-Sky Data Release \citep{Skrutskie06} to produce a sample of targets for spectroscopic follow-up. A total of 37 stars with $K_S < 4.8$, $J-K_S > 1.1$ were selected in the rectangle defined by the galactic coordinates $13^\circ  \lesssim l \lesssim 16^\circ$, $0^\circ  \lesssim b \lesssim 2^\circ$. The $K_S$ magnitude limit corresponds to a red supergiant with absolute magnitude $M_K = -7.7$ at a distance modulus of 11.5~magnitudes, obscured by a foreground extinction $A_K = 1$~mag. The $J-K_S$ color limit ensures that also red supergiants with virtually no foreground extinction are included in our sample. The sample is expected to be considerably contaminated by foreground red giant branch (RGB) stars, asymptotic giant branch (AGB) stars, and perhaps other types of cool stars, thus requiring spectral classification to confirm the presence of red supergiants.

\section{Observations and spectral classification\label{observations}}

We obtained spectra of the 37 stars in our sample using CAFOS, the facility imager and low-resolution spectrograph at the 2.2m telescope of the Calar Alto Observatory, on the nights of 22-23 and 23-24 June 2018. A single spectroscopic configuration was used consisting of a grism providing a coverage over the $5900-9000$~\AA \ range and a slit width of 1''5, yielding a resolution $R = 1300$. Exposure times were adjusted according to the red magnitude of the target, ranging from 120s to 1800s. The spectra were reduced and extracted using standard IRAF tasks, and calibrated in wavelength using the internal Hg, He and Rb arc lamps in the calibration unit of the instrument.

\begin{table*}[t]
\caption{Grid of MK standards used for classification}
\begin{tabular}{cccccc}
\hline
       & \multicolumn{5}{c}{Luminosity class} \\
       & Ia    & Iab      & Ib        & II & III \\
\hline
\noalign{\smallskip}
K2 &           &          & HD 160371 & HD 157999 & HD 173780 \\
K5 &           &          & HD 216946 &           &           \\
K7 &           &          &           & HD 181475 & HD 194193 \\
M0 &           &          &           &           & HD 6860   \\
M1 & HD 339034 &          & HD 163755 &           & HD 204724 \\
M2 & HD 239978 & HD 13136 & HD 10465  &           & HD 219734 \\
M3 &           &          &           & HD 40239  & HD 224427 \\
M4 &           &          &           & HD 175588 & HD 4408   \\
\hline
\end{tabular}
\label{std}
\end{table*}

Spectral classification was carried out by using a grid of MK standards obtained from the catalog of \citet{Garcia89} covering spectral types K2 to M4 (the latter corresponding to the temperatures of the coolest red supergiants) and luminosity classes Ia to III, which were observed with an identical instrument setup and reduced in the same way as our target observations. The grid of comparison standards is presented in Table~\ref{std}. Many of our targets turned out to be very late-type giants with temperatures and spectral types below the range covered by red supergiants, which we classified for the sake of completeness using the atlas presented in Figure~7 of \citet{Fluks94}.

\begin{figure}[ht]
\begin{center}
\hspace{-0.5cm}
\includegraphics [width=9cm, angle={0}]{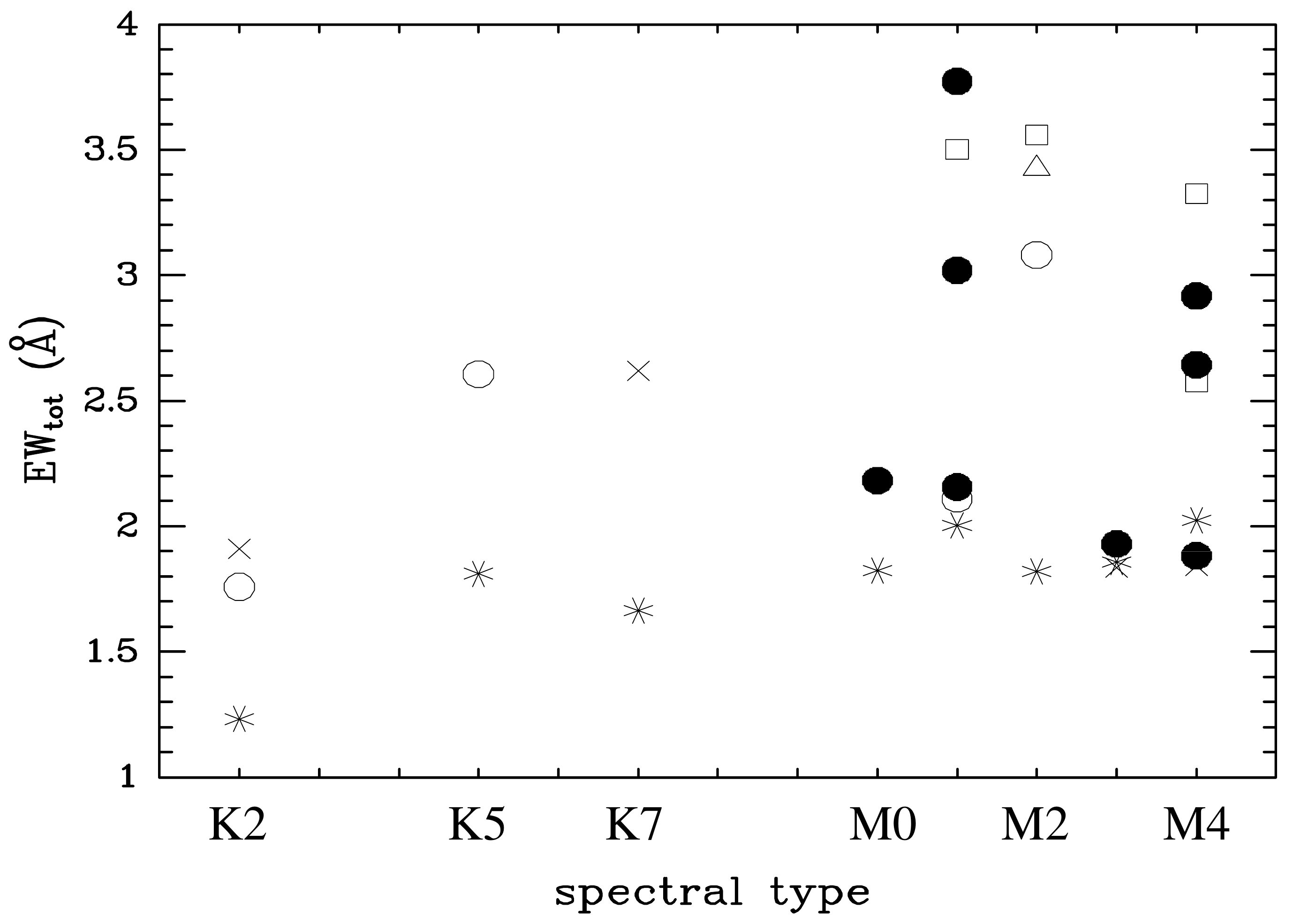}
\caption []{Sum of the equivalent widths of the five luminosity-sensitive spectral features noted in Table~\ref{eqw}, both for the MK standards listed in Table~\ref{std} and for our targets stars with spectral types M4 or earlier. Squares represent luminosity class Ia standards; triangles, class Iab; open circles, class Ib; crosses, class II; and asterisks, class III. The filled circles correspond to our targets. We consider the four stars with a sum of equivalent widths above 2.5 as {\it bona fide} supergiants, as no class III standard populates that region.}
\label{fig_eqw}
\end{center}
\end{figure}

\begin{figure}[ht]
\begin{center}
\hspace{-0.5cm}
\includegraphics [width=9cm, angle={0}]{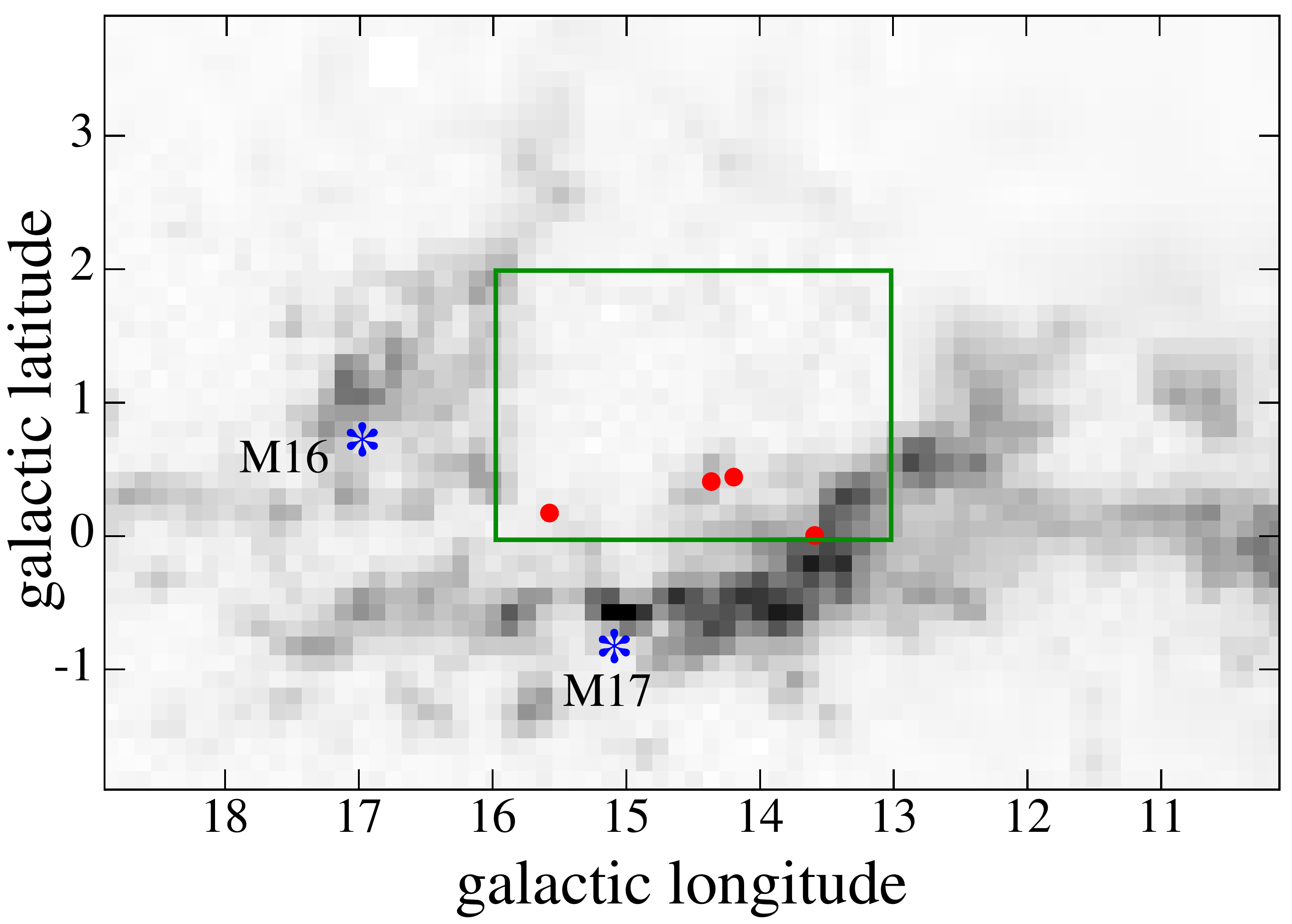}
\caption []{Map CO emission in the velocity range $+15 \lesssim v_{\rm LSR} {\rm (km \ s^{-1})} \lesssim +20$, showing the open shell that dominates the distribution between $13^\circ < l < 18^\circ$ and the positions of the giant HII regions M16 and M17 superimposed on its rim. The rectangle marks the area where we selected red supergiant candidates, and the four circles show the locations of the four confirmed red supergiants that we find.}
\label{map}
\end{center}
\end{figure}

The differences between red supergiants and luminosity class III giants in the spectral range K0-M4 in the wavelength region covered by our spectra are rather subtle. The problem has been studied in detail by \citet{Dorda16,Dorda18} using principal component analysis on large samples of luminous red stars. For the spectral range and resolution of our spectra, a convenient and highly reliable criterion is based on selected surface gravity-sensitive lines in the Calcium triplet region, already noted by \citet{Keenan45}. We used for this purpose the KI lines at 7665~\AA\ and 7699~\AA\ and the FeI lines at 8468~\AA\ (blended with Ti), 8514~\AA\, and 8469~\AA , all of which have a positive luminosity class effect. We measured their equivalent widths in the spectra of both our targets and the MK standards, and used the sum of equivalent widths of all five features to identify the red supergiants in our sample. Figure~\ref{fig_eqw} shows the separation of the MK standards between giants and supergiants reflected in the sum of equivalent widths and the positions of our targets in the same plot, where we can see that four of the stars, J18143590-1621473, J18145920-1707138, J18145951-1614573, and J18182574-1518131, appear well into the region occupied by red supergiants. The measured equivalent widths are given in Table~\ref{eqw}, together with the spectral classifications of the whole sample, and their locations are shown in Figure~\ref{map}.

\begin{table*}[t]
\caption{Spectral classification and equivalent widths of luminosity-sensitive features}
\begin{tabular}{lccccccc}
\hline
     &         & \multicolumn{6}{c}{Equivalent widths (\AA )} \\
Star & Sp. type & KI   & KI    & FeI+TiI & FeI  & FeI  & Sum \\
     &          & 7665 & 7699  & 8468    & 8514 & 8689 &     \\
\hline
\noalign{\smallskip}
J18065203-1642331 & M8   & & & & & & \\
J18074880-1624225 & M6   & & & & & & \\
J18092210-1617158 & M6.5 & & & & & & \\
J18101196-1657356 & M5   & & & & & & \\
J18104630-1543262 & M6   & & & & & & \\
J18110252-1608253 & M6   & & & & & & \\
J18111093-1607391 & M9.5 & & & & & & \\
J18112913-1550149 & M6.5 & & & & & & \\
J18115512-1453007 & M7   & & & & & & \\
J18121334-1416310 & M6   & & & & & & \\
J18121448-1523575 & M7   & & & & & & \\
J18124479-1630225 & M5   & & & & & & \\
J18132372-1503002 & M5   & & & & & & \\
J18133539-1658565 & M6   & & & & & & \\
J18140672-1545547 & M7   & & & & & & \\
J18142432-1707379 & M0   & 0.416 & 0.193 & 0.606 & 0.534 & 0.433 & 2.182 \\
J18143478-1500562 & M4   & & & & & & \\
J18143590-1621473 & M1   & 0.735 & 0.510 & 1.051 & 0.871 & 0.605 & 3.772 \\
J18143895-1700169 & M7   & & & & & & \\
J18144293-1627321 & M10  & & & & & & \\
J18144694-1436499 & M6.5 & & & & & & \\
J18144948-1445478 & M6   & & & & & & \\
J18145920-1707138 & M4   & 0.591 & 0.296 & 0.788 & 0.532 & 0.436 & 2.643 \\
J18145951-1614573 & M1   & 0.597 & 0.478 & 0.776 & 0.611 & 0.557 & 3.019 \\
J18151511-1605135 & M6   & & & & & & \\
J18151718-1615082 & M6.5 & & & & & & \\
J18152673-1442534 & M6   & & & & & & \\
J18152765-1630511 & M4   & 0.333 & 0.152 & 0.573 & 0.436 & 0.388 & 1.882 \\
J18153653-1614197 & M1   & 0.319 & 0.185 & 0.589 & 0.559 & 0.504 & 2.156 \\
J18155478-1535526 & M7   & & & & & & \\
J18155937-1628566 & C    & & & & & & \\
J18161687-1546501 & M6   & & & & & & \\
J18162315-1625103 & M3   & 0.322 & 0.084 & 0.609 & 0.515 & 0.399 & 1.929 \\
J18164708-1533279 & M5   & & & & & & \\
J18165624-1549457 & M6   & & & & & & \\
J18182574-1518131 & M4   & 0.520 & 0.346 & 0.803 & 0.661 & 0.587 & 2.917 \\
J18183294-1524032 & M6.5 & & & & & & \\
\hline
\end{tabular}
\label{eqw}
\end{table*}

The trigonometric parallax measurements of J18145951-1614573 and J18182574-1518131 are listed respectively as $0.4751 \pm 0.1689$~mas and $0.4935+/-0.1044$~mas respectively in the Gaia DR2 catalog \citep{GaiaDR218}, in excellent agreement with the estimated distances to M16 and M17 quoted in Section~\ref{targets}. A third star, J18145920-1707138, also has a parallax consistent with that distance, $0.7037 \pm 0.2863$~mas, well within its error bar. The parallax of the fourth star, J18143590-1621473, is listed as $0.1472+/-0.0951$~mas, suggesting a much farther distance of over 4~kpc, or even 6.8~kpc at the central value of 0.1472~mas. The temperature and luminosity implied by such distance is still within the range allowed by evolutionary models that we discuss in Section~\ref{parameters}, but we notice that this star is within only $8'9$ of J18145951-1614573, one of the two stars whose parallax best matches the distance to M16 and M17, and has an even lower amount of foreground extinction. We thus consider it more likely that it does belong to the complex  under study, rather than being a background supergiant.

Two other stars, J18142432-1707379 and J18153653-1614197, have a sum of equivalent widths marginally above the narrow band defined by giants (see Table~\ref{eqw}). This region is also populated by a few class I and II giants. To discern their true nature, we note that the parallax measured by Gaia for J18153653-1614197 is $1.0153 \pm 0.0653$, thus strongly suggesting that it is a foreground red giant. Concerning J18142432-1707379, the Gaia DR2 catalog lists a negative parallax ($-0.3415 \pm 0.3103$), suggesting problems with the measurements. We do not consider either star among our sample of red supergiants.

\section{Results\label{results}}

\subsection{Physical parameters\label{parameters}}

\begin{figure}[ht]
\begin{center}
\hspace{-0.5cm}
\includegraphics [width=9cm, angle={0}]{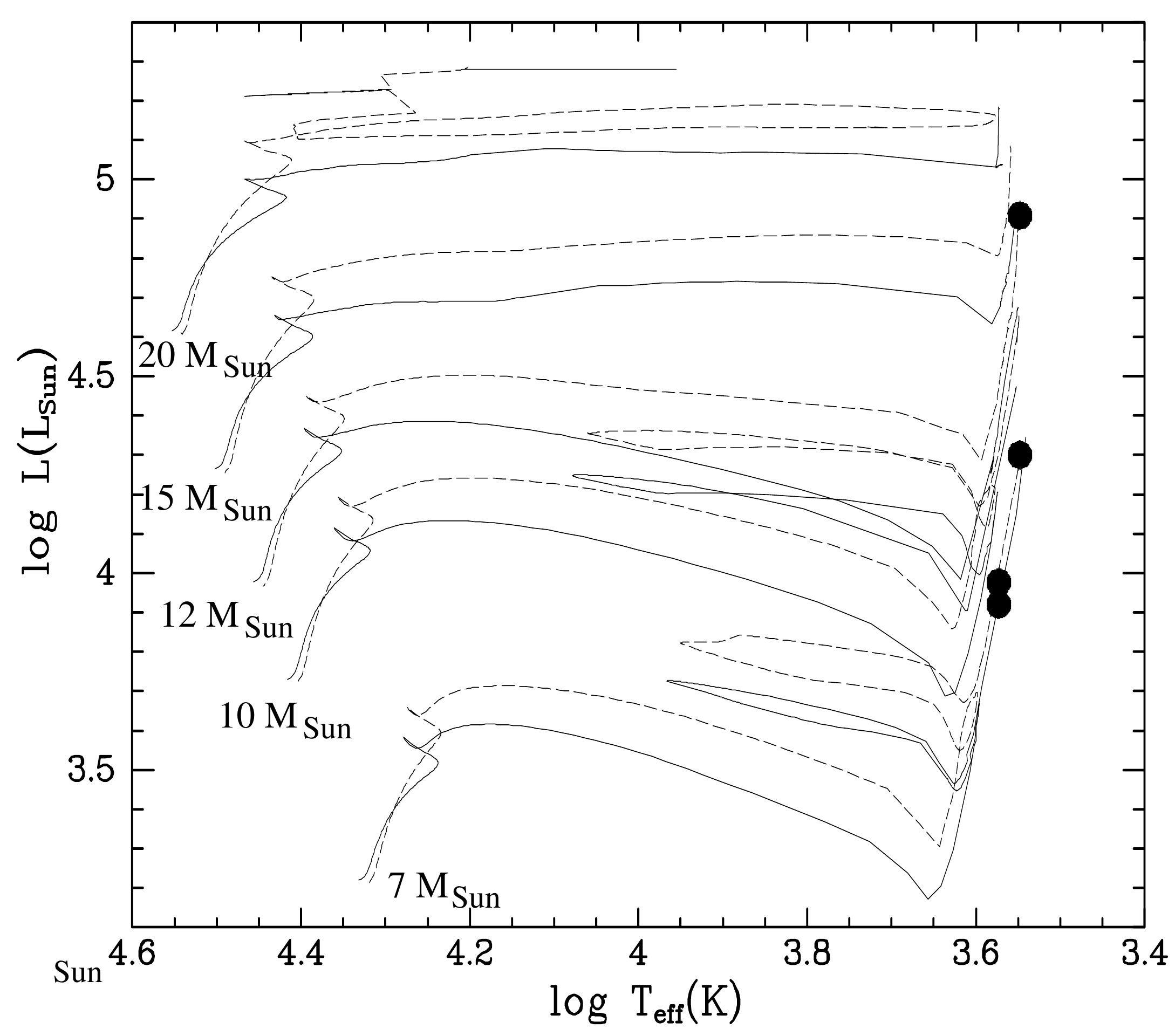}
\caption []{Position of the four red supergiants on the temperature-luminosity diagram. The Geneva evolutionary tracks \citep{Ekstroem12} are plotted for stars with different initial masses, both non-rotating (solid lines) and rotating with a velocity $40$~\% of the critical velocity.}\label{TeffL}
\end{center}
\end{figure}

We used the intrinsic colors of galactic red supergiants of \citet{Elias85} to estimate the extinction and absolute magnitude $M_K$ of all four red supergiants, assuming a common distance modulus for all of them. Temperatures and bolometric corrections were derived using the calibrations of \citet{Levesque05} and \citet{Levesque07} for galactic red supergiants. The extinction estimate was made from the $(V-K_S)$ color using the 2MASS $K_S$ magnitude and the $V$ magnitude listed in the NOMAD catalog \citep{Zacharias04}. The positions of the stars in the temperature-luminosity diagram were then compared with the Geneva evolutionary tracks for massive stars with and without rotation \citep{Ekstroem12}, as shown in Figure~\ref{TeffL}. Initial masses and ages were then derived by assuming that all the stars are on the steady evolutionary part of its evolutionary track, corresponding to the phase of helium burning at the core, before the upturn immediately preceding the supernova explosion or, for rotating stars, before the blue loop excursion to higher temperatures that follows the end of He burning. Although taken at
face value the temperatures adopted for the stars places them right on the upturn, the very short duration of that
stage leads us to believe that this is almost surely due to a mismatch between the temperatures predicted by the evolutionary tracks and the temperature calibration, in the sense of the latter being colder. The difficulties and uncertainties in the determination of galactic red supergiant temperatures have been discussed by \citet{Levesque05}.

\begin{table*}[t]
\caption{Parameters of red supergiants}
\begin{tabular}{lcccccccccccc}
\hline
                  &         &        &     &       &       &       &          &  & \multicolumn{2}{c}{Non-rotating} & \multicolumn{2}{c}{Rotating ($0.4 v_{\rm crit}$)} \\
Star              & $l$     & $b$    & $V$ & $K_S$ & $A_K$ & $M_K$ & $\log L$ & $\log T_{\rm eff}$ & Mass & Age & Mass & Age \\
                  & $(^\circ)$ & $(^\circ)$ &  &     &       &     &  $({\rm L}_\odot)$ & (K) & (M$_\odot$) & (Myr) & (M$_\odot$) & (Myr) \\

\hline
\noalign{\smallskip}
J18143590-1621473 & 14.223 & 0.471 & 11.45 & 4.15 & 0.43 & -7.78 & 3.920 & 3.573 & 10 & 22.3-24 & 9  & 28-30 \\
J18145920-1707138 & 13.602 & 0.027 & 12.20 & 3.09 & 0.49 & -8.90 & 4.300 & 3.548 & 12 & 15-17   & 11 & 21-23 \\
J18145951-1614573 & 14.368 & 0.442 & 15.05 & 4.43 & 0.85 & -7.92 & 3.976 & 3.573 & 10 & 22.3-24 & 9  & 28-30 \\
J18182574-1518131 & 15.593 & 0.164 &  8.91 & 1.36 & 0.29 &-10.42 & 4.908 & 3.548 & 18 & 9.5-10  & 15 & 14-15 \\
\hline
\end{tabular}
\label{physpar}
\end{table*}

Table~\ref{physpar} lists the derived physical parameters of each star, for the two cases of a non-rotating precursor and of a precursor rotating at $0.4$ times the critical velocity. Rotation produces a moderate increase of the luminosity for a star of a given mass, a longer duration of the main sequence phase, and a slightly longer duration of the He burning phase. The presence in the region of supergiants with widely different luminosities, and therefore widely different ages, implies an extended star forming period.

\subsection{Estimating the number of massive stars and supernovae\label{obsn}}

Given the short duration of the red supergiant phase, the fact that we can observe simultaneously four stars undergoing it simultaneously hints at a large number of precursors during the star formation history of the region. We can use the number of currently observed M supergiants, together with simple assumptions on the star formation rate and a standard initial mass function, to obtain a crude estimate of the number of massive stars that have existed in the region and the number of core-collapse supernovae that have happened, leading to an estimate of the mechanical energy dumped into the bubble.

The expected number of red supergiants $N_{\rm RSG}$ observed at present ($t=0$) can be expressed as

\begin{equation}
\label{nrsg}
%\begin{split}
N_{\rm RSG} = \int_{t=-\infty}^0 \int_{M = 0}^\infty \Psi(t) V_{\rm RSG}(M,t) A \xi(M) dM dt
%\end{split}
\end{equation}

\noindent where $\Psi(t)$ is the star formation rate, $\xi(M)$ is the initial mass function, $A$ is a normalization factor, and $V_{\rm RSG}(M,t)$ is a visibility function: $V_{\rm RSG} (M, t) = 1$ if a star of mass $M$ formed at time $t$ is at the red supergiant phase at present, and zero otherwise \footnote{An equivalent treatment is presented in \citet{Comeron16}. For simplicity we have removed here the integration over the rotation velocity and we consider the two cases of non-rotating and rotating precursors, listed in Table~\ref{physpar}.}

We make the rough assumption that $\Psi(t)$ was constant during the whole period in which the stars currently observed as red supergiants were born. We further assume, rather conservatively, that the star formation history of the region extends from the time of formation of the least massive red supergiant observed at present, until the time when the most massive red supergiant now visible formed. Assuming that the star formation history is limited in this way makes the limits of the integration over time and mass largely irrelevant, as the expression within the integral has a value of zero for masses below that of the least massive red supergiant, and also for masses above that of the most massive one. Adopting a power law with the Salpeter exponent $-2.35$ for the high-mass end of the initial mass function, $\xi(M) = M^{-2.35}$, the current observation of $N_{\rm RSG} = 4$ sets the value of the normalization factor $A$.

The number of massive stars down to a mass $M_{\rm lim}$ that have formed in the region, which we denote as $N_{\rm OB}$ for convenience (although $M_{\rm lim}$ does not necessarily have to correspond to the mass of a B star), can be expressed in terms of the quantities introduced in the previous equation:

\begin{equation}
\label{nob}
%\begin{split}
N_{\rm OB} = \int_{t=-\infty}^0 \int_{M = M_{\rm lim}}^\infty \Psi(t) A \xi(M) dM dt
%\end{split}
\end{equation}

Using Eq.~\ref{nrsg},

\begin{equation}
\label{nobnorm}
%\begin{split}
N_{\rm OB} = {{\int_{t=-\infty}^0 \int_{M = M_{\rm lim}}^\infty \Psi(t) \xi(M) dM dt} \over
{\int_{t=-\infty}^0 \int_{M = 0}^\infty \Psi(t) V_{\rm RSG}(M,t) \xi(M) dM dt}} N_{\rm RSG}
%\end{split}
\end{equation}

Calculated in this way $N_{\rm OB}$ includes all the stars down to mass $M_{\rm lim}$ that have formed in the region, regardless of whether the star still exists or has exploded already as supernova. If $M_{\rm lim}$ is chosen such that the lifetime of a star of that mass is shorter than the time in the past when star formation in the region ceased, no stars with a mass higher than $M_{\rm lim}$ should be observable at present in the region. To estimate the total number of supernovae $N_{\rm SN}$ a similar expression can be written:

\begin{equation}
\label{nsnnorm}
%\begin{split}
N_{\rm SN} = {{\int_{t=-\infty}^0 \int_{M = M_{\rm SN}}^\infty \Psi(t) V_{\rm SN} (M,t) \xi(M) dM dt} \over
{\int_{t=-\infty}^0 \int_{M = 0}^\infty \Psi(t) V_{\rm RSG}(M,t) \xi(M) dM dt}} N_{\rm RSG}
%\end{split}
\end{equation}

\noindent where $M_{\rm SN}$ is the minimal initial mass for a star to end its life as a core collapse supernova, for which we adopt $M = 8$~M$_\odot$. $V_{\rm SN} (M,t)$ has a value of zero if the absolute value of $t$ is less than the lifetime of a star of mass $M$, and unity otherwise.

Using $N_{\rm RSG} = 4$ and the ages listed in Table~\ref{physpar}, we estimate the number of O-type stars that existed in the history of the association (down to $M = 17.5$~M$_\odot$; \citet{Pecaut13}) to be between 29 (adopting ages from models with no rotation) and 37 (rotation with $v=0.4 v_{\rm crit}$). The number of core collapse supernovae is estimated to be between $N_{\rm SN} \sim 82$ (no rotation) and 111 (with rotation). Given the old age of the lightest red supergiants currently observed in the region, we expect that a large number of stars with early B-type precursors ($M < 17.5$~M$_\odot$) have already exploded as supernova.

\subsection{Bubble energetics \label{energy}}

A similar computation allows us to estimate the energy dumped inside the bubble blown around the ancient association by OB stars and supernovae. The kinetic energy deposited inside the bubble by stellar winds is

\begin{equation}
\label{energy_tot}
%\begin{split}
E_{\rm wind} = {{{\int_{t=-\infty}^0 \int_{M = 0}}^\infty \Psi(t) L_{\rm w}(M,t) \tau(M) \xi(M) dM dt} \over
{\int_{t=-\infty}^0 \int_{M = 0}^\infty \Psi(t) V_{\rm RSG}(M,t) \xi(M) dM dt}} N_{\rm RSG}
%\end{split}
\end{equation}

\noindent where $L_{\rm w}(M) = 0.5 {\dot M} v_\infty^2$ is the mechanical luminosity of the wind of a star of mass $M$ having a mass loss rate ${\dot M}$ and a terminal wind velocity $v_\infty$. $\tau(M)$ is the time over which the star has been producing the wind: $\tau(M) = {\rm min}(-t,\tau_{\rm OB} (M))$, where $\tau_{\rm OB} (M)$ is the main sequence lifetime of a OB star of mass $M$. We neglect the short-lived Wolf-Rayet and blue supergiant phases as their overall contribution to the energetics of the bubble is comparatively small, as well as the red supergiant phase in which the slow wind makes the contribution to the mechanical luminosity integrated over time irrelevant. We have used the compilation of data of OB stars with normal winds by \citet{Martins05} to derive empirical $\log {\dot M}$-$\log L$ and $\log {\dot M} v_\infty \sqrt{R}$-$\log L$ relations, where $L$ is the luminosity of the star and $R$ is its radius, which we derive from the $\log L$ and $\log T_{\rm eff}$ values for stars of different masses compiled by \citet{Pecaut13}.

The integration of Eq.~\ref{energy_tot} leads us to estimate $E_{\rm wind} \sim 10^{52}$~erg. The decrease of both ${\dot M}$ and $v_\infty$ with decreasing stellar mass implies that $E_{\rm wind}$ is dominated by high-mass stars, in spite of the declining initial mass function and main sequence lifetimes with increasing mass. Under our assumption that massive star formation in the region ended at the time when the most massive red supergiant currently observed was born, the contribution of stellar winds to the bubble energetics ceased to be important long ago, as the most massive stars with the most powerful winds ended their main sequence life. On the other hand, stars with lower masses formed over the history of the region have continued to explode as core-collapse supernovae until the present, and currently dominate the rate of injection of energy into the bubble. Adopting a kinetic energy of $10^{51}$~erg per supernova, the numbers given in Sect.~\ref{obsn} suggest an energy input of $\sim 10^{53}$~erg due to supernovae.

Within the rough approximations and extrapolations made, the energy dumped by stellar winds and (mainly) supernovae in the interior of the bubble appears to be sufficient for it to have rearranged the molecular gas in the region over the large region delineated by the contour of the CO shell. If the bubble were powered by the energy deposited at a roughly continuous rate over a period of 30~Myr and expanding in a uniform medium of density $n_0$, its radius at present would be

\begin{equation}
\label{r_bubble}
%\begin{split}
R = 1.65 \times 10^{-6} \bigl({{L_{\rm m} ({\rm erg \ s^{-1}})} \over {n_0 ({\rm cm}^{-3})}} \bigr)^{1/5} t ({\rm Myr})^{3/5}
%\end{split}
\end{equation}

\noindent \citep[e.g.][and references therein]{Bisnovatyi95}, where $L_{\rm m}$ is the kinetic energy injected in the bubble per unit time, which we approach as $L_{\rm m} \sim 10^{53} {\rm erg} / 30 \ {\rm Myr} \sim 10^{38} {\rm erg \ s}^{-1}$ and $t$ is the current age of the bubble, which we take as 30~Myr. If the original association formed in a giant molecular cloud with the typical average density $n_0 \sim 100$~cm$^{-3}$, the expected radius of the bubble would be of the order of $R \sim 200$~pc. This is significantly larger than the actually measured $\sim 70$~pc, but the discrepancy should not be taken as ruling out the interpretation of the shell as the edge of the bubble blown by the massive members of the old association: the stellar and supernova content of the association and the duration and history of the star forming activity are two of the main unknowns where we have made educated guesses rather than accurate reconstructions. The history of star formation, and therefore of the injection of mechanical energy, can strongly influence the pattern of expansion of a bubble \citep{Silich96}. Furthermore, the radius derived from Eq.~\ref{r_bubble} exceeds the vertical scale height of the HI layer of the galactic disk \citep{Dickey90}, and the effect of evolution in a density-stratified medium is obvious in the shape of the shell, which strongly suggests that the bubble has blown up and has vented much of the energy into the galactic halo. Concerning the expansion velocity of the shell, $v = d R / d t$, the adopted quantities yield $v \simeq (10 \ n_0^{1/5} {\rm cm}^{-3})$~km~s$^{-1}$, close to the sound speed in the warm interstellar medium, implying that the expansion of the shell still causes a strong compression of the swept-up gas.

\section{Discussion and conclusions \label{discussion}}

We have developed the interpretation of the presence of four red supergiants in the interior of the shell in whose rim M16 and M17 are located as an indication for the past existence of a rich OB association in the region. The association, which formed massive stars over an extended period lasting from $\sim 30$~Myr to $\sim 10$~Myr ago, may have hosted several tens of O stars and the precursors of around one hundred supernovae up to the present time. Such an extended period of star formation may not be an anomalous feature of large, rich OB associations, based on similar recent findings on the younger, more nearby Cygnus OB2 association \citep{Hanson03,Comeron12,Comeron16,Berlanas18}.

Our estimates about the richness of the association and its supernova history involve some grossly oversimplifying assumptions, and should not be taken as a quantitative proof of the hypothesis. However, we have shown that the injection of mechanical energy into the bubble by the postulated association could have produced a substantial rearrangement of the gas over hundred parsec-long scales, accounting for the current size of the shell, and an expansion at supersonic speeds up to the present, thus being able to compress the swept-up gas or previously existing  gas concentrations encountered along the path of expansion, ultimately triggering further star formation. The origin and location of M16 and M17, two giant HII regions much younger than the estimated age of the old association, as a consequence of triggered star formation along the rim of the shell thus becomes a plausible scenario. A more realistic modeling should take into account at least a time-variable energy input consistent with the assumed star formation rate of the association and the expansion of the shell in a vertically stratified medium.

Proposed examples of triggered star formation abound in the literature over a variety of scales. In particular, the high quality infrared observations from the {\sl Spitzer} and {\sl Herschel} space observatories over the last decade and a half have provided abundant support for the actual existence of triggered star formation along the edges of bubble-like HII regions \citep{Deharveng11, Samal14,Liu16,Gama16,Cichowolski18}. Examples at larger scales where star formation appears to have been caused by expanding supershells seen in HI in the disk of our Galaxy also exist \citep{Megeath03,Oey05,Arnal07,Lee09}, although the evidence is more elusive, largely due to confusion and crowdedness along the line of sight. Cleaner examples exist in the Magellanic Clouds \citep{Oey95,Efremov98} and other nearby galaxies with resolved populations \citep{Egorov14,Egorov17}, where the evidence for large-scale triggered star formation is more compelling.

The hints reported here of an old association having caused the large molecular shell and possibly having triggered the formation of M16 and M17 along its rim, based on just four red supergiant stars, make it difficult to further confirm the proposed scenario. A prediction that may be validated in the near future is the existence of large amounts of early B-type stars, the still unevolved lower-mass counterparts of the red supergiant precursors, near the main sequence turn-off of the association. The location of those stars near the galactic equator makes it nearly impossible to identify them at present against the overwhelming foreground and background contamination. However, the final Gaia data release, with parallaxes of substantially better quality than the already excellent ones available at present, may make it possible to produce samples within precise distance boundaries in which early-type stars could be easily distinguished. The coincidence at a similar distance of two young giant HII regions, an evolved supershell, and the remnants of an old, rich association in its interior might not be the ultimate proof of the triggered common origin of M16 and M17, but would strongly support it.

\begin{acknowledgements}

We are very pleased to thank once again the staff at the Calar Alto Observatory, especially on this occasion by
Gilles Bergond and David Galad\'\i . FC thanks the hospitality of the University of Barcelona during earlier phases of this work. The Two Micron All Sky Survey (2MASS) is a joint project of the University of Massachusetts and the Infrared Processing and Analysis Center/California Institute of Technology, funded by the National Aeronautics and Space Administration and the National Science Foundation. This research has made use of the SIMBAD database, operated at CDS, Strasbourg, France.

\end{acknowledgements}

\bibliographystyle{aa} % style aa.bst
\bibliography{msg_M16M17_cit}

%---------------------------------------------------------
% Edit the file globule_ref.bib to update the references.
% Then run the commands:
%       latex globule
%       bibtex globule
%----------------------------------------------------------

%\bibliography{globule_ref}

%\Online

%\listofobjects

\end{document}